# *Finite Population Sampling as $n \rightarrow N$:*
# *Empirical Evidence for the Transition from Inference to Accuracy*

Dr. Mike Crowhurst
Graduate Student
Dr. Bing Zhang Department of Statistics
University of Kentucky

***Abstract***

The Central Limit Theorem provides a foundation for inferential statistics and hypothesis testing. It describes how standardized statistics behave under repeated sampling from large populations. However, if the size of the sample ($n$) becomes so large that it approaches the size of the population ($N$), sampling variability becomes so small that standard error and margin of error both approach zero. The purpose of this project was to investigate the behavior of estimators such as the sampling fraction ($f = {}^{n}/_{N}$ $as\ f \rightarrow 1$), motivated by modern data streams from administrative records, transaction logs, sensor streams, and institutional databases that capture large parts of finite populations. We constructed two finite populations with known parameters and took repeated samples across a range of sampling fractions. We then examined the resulting randomization distributions of the sample mean to understand the collapse of sampling variability. We also conducted additional experiments using various CPU- and GPU-based methods to evaluate the deviation of the sample mean from the defined population mean under different conditions. The results confirm that sampling variability does diminish in as expected under the finite population theory and it becomes negligible well before full enumeration is reached. Once sampling variability is minimized, remaining deviation in estimators relates more to numerical precision and computational structure, rather than by random sampling. These findings support a reassessment of inferential assumptions in high-coverage, large-scale settings.

***Introduction***

The Central Limit Theorem (CLT) (de Moivre, 1733) (Laplace, 1812) (Lyapunov, 1901) (Feller, 1971) in inferential statistics tells us that if you repeatedly draw many independent samples of size $n$ from the same population, then the distribution of a properly standardized statistic (like the sample mean) approaches a normal distribution. That is true even if the population is not normally distributed. But the assumption behind CLT is that we are dealing with a relatively small sample from a large population.

What if the sample is so large that the unsampled portion of the population is negligible?

The CLT becomes less applicable as the sampling fraction ($n/N$) approaches one, because sampling variability is mechanically reduced by the finite population correction leaving little uncertainty for the asymptotic approximation to characterize. In a sense the "bell curve" that is used to show a randomization distribution becomes so narrow that it ultimately becomes a vertical line (see Figure 1). There is no distribution left, only a constant:

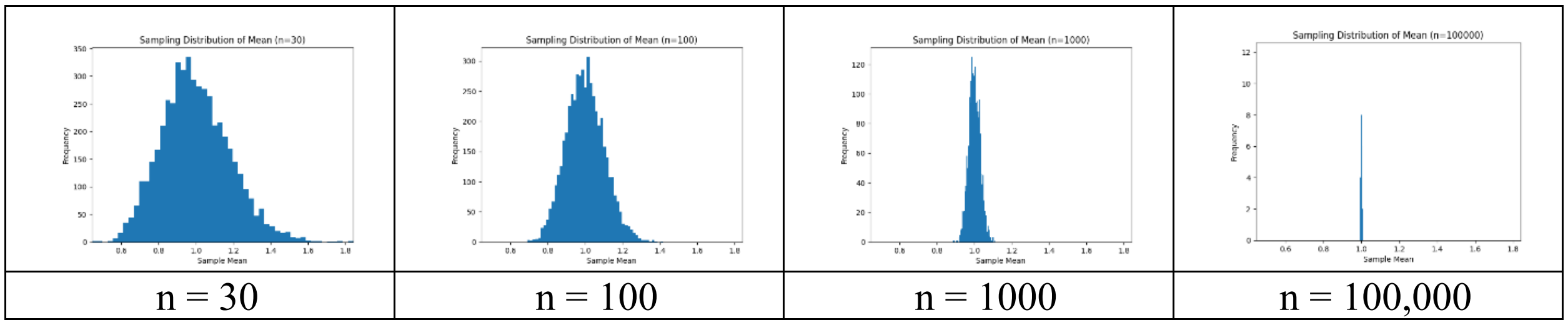

***Figure 1: Randomization distributions with n's of 30, 100, 1000, and 100,000 (generated by A.I.)***

This question is no longer purely theoretical as advances in data collection, storage, and computational capacity have led to widespread use of systems that generate near-enumeration data rather than small random samples. Cochran (Cochran, 1977) and others (Lohr, 2021) (Sarmda & . C.-E., 1992) note that:

$$Var(\bar{y}) \approx \left(1 - \frac{n}{N}\right) \cdot \frac{s^2}{n}$$

The variance decrease when the sampling fraction becomes zero (when $n \approx N$). Lohr (Lohr, 2021) notes that when $n/N$ is small the Finite Population Correction (FPC) is small and further, that when $n/N$ is large the notion of independence breaks down.

Data streams from administrative records, transaction logs, sensor streams, and institutional databases increasingly capture large fractions of finite populations, sometimes approaching full coverage. As the sampling fraction approaches one, $1 - n/N$ (the FPC factor) drives the variance to almost zero. Nevertheless, much statistical practice continues to rely on inferential frameworks grounded in independent and identically distributed (i.i.d.) or superpopulation models, in which sampling variability is treated as the primary source of uncertainty about unknown parameters, even in settings where data exhibit near-census coverage or non-random selection. This is an argument between classical inference (frequentist) thinking and a modern data reality that 1) data may be nearly complete, and 2) uncertainty is a combination of data generation, selection, and measurement.

When the observed fraction of a finite population becomes large, this interpretation becomes problematic: sampling variability collapses mechanically as the population is exhausted, even though analysts may continue to compute standard errors, confidence intervals, and normal approximations. Understanding how and when classical inferential reasoning breaks down in this near-enumeration regime is therefore essential for interpreting modern data analyses.

The purpose of this project was to study the transition between two regimes. First, there is that of the CLT and its related reasoning. Second, there is the near-enumeration regime where the population is finite, sampling is without replacement, and $n/N \to 1$.

To visualize what this second regime looks like, consider the behavior of repeated samples drawn without replacement from a finite population as the sampling fraction increases. In contrast to the CLT regime, where observations are approximately independent, sampling without replacement induces increasingly strong dependence among observations as $n/N$grows. As a result, sampling variability collapses mechanically: the variance of common estimators approaches zero, and repeated samples become increasingly similar. When the sample size ($n$) equals the population size ($N$), the distinction between sample and population disappears entirely, and there is no randomness. This enumeration regime is therefore characterized not by improved inference, but by the exhaustion of random variation through population coverage. It is a situation that arises naturally in modern data systems that observe large fractions of finite populations. In this regime, statistical analysis shifts from inference toward accuracy analysis, error decomposition, explicit modeling of measurement and processing error, and consideration of numerical stability and algorithmic bias (Meng, 2018), (Groves, 2011), (Lohr, 2021).

In this project we will use a GPU-based simulation and experimentation platform to study this transition empirically. Across these experiments, we systematically vary population distributions, estimator formulations, and numerical precision in order to isolate how each factor influences the collapse of sampling variability as the sampling fraction increases.

The computational approach allows us to move beyond symbolic results such as the finite population correction and instead examine the *rate* at which sampling variability collapses in practice. We study how the dispersion of classical estimators diminishes, when normal approximations cease to be informative, and when repeated samples become nearly identical. In this near-enumeration regime, statistical uncertainty arising from sampling is no longer the dominant source of error.

Our GPU-based implementation is designed around a simple but scalable structure: population generation, parallel random sampling without replacement, and parallel computation of estimators across large numbers of hypothetical samples. The use of GPUs is essential, as the phenomena of interest become visible only when both population sizes and the number of repeated samples are large. These scales make it possible to resolve extremely small sampling variances and to compare them meaningfully against numerical effects such as floating-point rounding, accumulation error, and algorithmic bias.

In this way, the project shifts emphasis from classical inference based on the Central Limit Theorem toward an accuracy-oriented perspective appropriate for near-enumeration data. When sampling variance collapses (approaches zero), remaining uncertainty is driven by measurement error, processing decisions, and numerical stability rather than by random sampling. The results illustrate how statistical reasoning must change as data collection approaches full enumeration, and why large sample size alone is insufficient to guarantee reliable inference. Understanding the implications of this transition is therefore essential for interpreting statistical results derived from modern near-enumeration data.

**What we are doing differently**

In this work, we intentionally move away from traditional inferential statistics framework by separating the usual identification between sampling variability and uncertainty about an unknown population parameter. Conventional inference usually assumes an effectively infinite population, independent and identically distributed observations, and unknown parameters. In this context, the randomization distribution of a statistic serving as a substitution for uncertainty. Our approach begins with an clearly defined finite population whose properties are known exactly. The population mean and variance are defined by direct computation rather than inference. Randomness only comes into play through sampling without replacement from this fixed population. Here, the randomization distribution does not show uncertainty about an unknown truth; it instead highlights the remaining randomness caused by partial observation of a finite population. As sampling proceeds without replacement, the dependence among observations grows, and the variance of the sample mean is influenced by the finite population correction factor $(1-f)$, where $f = n/N$. The resulting drop in estimator variability as $f \to 1$ is structural rather than inferential. Therefore, we treat the randomization distribution not as a tool for inference, but as a subject of study that shows how sampling variability develops and ultimately fades as we approach full enumeration.

**Why this shift is necessary**

This change in thinking is necessary because the assumptions of classical inferential reasoning often conflict with modern data situations that observe large fractions of a finite population. When the sampling fraction is small, the infinite-population model is adequate. Dependence among observations is minor, and we can view sampling variability as uncertainty about unknown parameters. However, as the sampling fraction grows, the dependence caused by sampling without replacement becomes significant. Sampling variability decreases according to the finite population correction $(1-f)$. In situations close to near-enumeration, the randomization distribution narrows not because information is being gained about an unknown quantity, but because there is less population remaining to randomize. When we have full enumeration, the randomization distribution reduces to a single point, and standard errors, confidence intervals, and normal approximations lose their meaningful interpretation. By explicitly considering a finite population with known truths and analyzing how estimator variability decreases as the sampling fraction increases, this work reveals a structural change that is often hidden by reasoning based on the central limit theorem. This viewpoint supports the broader goal of focusing less on inference and more on accuracy analysis, error decomposition, and numerical stability in situations where sampling uncertainty is no longer the main source of variability.

***Methods***

This study investigated how estimator behavior changed under finite-population sampling as the sampling fraction approached enumeration (n/N = 1). Rather than working within an abstract infinite-population framework, we constructed explicit finite populations with known parameters and treat those parameters as fixed reference values throughout the analysis. In this setting, randomness entered only through controlled sampling or through numerical computation; no unknown parameters are inferred.

The work was carried out in a sequence four phases. Each phase was designed to isolate a specific aspect of the transition from classical sampling behavior to near-enumeration. Phases 1–3 focus on the collapse of sampling variability, while Phase 4 what remains once that variability has effectively disappeared.

In phase 1, we constructed finite populations of a fixed size $N$. For each population, the mean $\mu$ and variance $\sigma^2$ were computed directly and treated as exact quantities rather than objects of inference.

We considered multiple population types to ensure the results were not tied to any particular distribution. These included well-behaved distributions (e.g., Normal) as well as more challenging cases (e.g., heavy-tailed Student-t). By doing this, we were able to observe whether the transition toward near-enumeration varied across different distributions.

In phase 2, we introduced randomness through sampling without replacement using simple random sampling. The key parameter governing this experiment was the sampling fraction,

$$f = \frac{n}{N}$$

which determines the proportion of the population to be observed.

For each value of $f$, we drew samples of size $n = \lfloor fN \rfloor$ while holding the population fixed. The purpose of this phase was to established the finite-population structure and to emphasize a basic but important point: as long as $f < 1$, the sample mean remains a random quantity, even though the population parameters are known exactly.

Phase 3 extended the sampling procedure by repeating the Phase 2 experiment many times at fixed values of the sampling fraction ($f$). These repetitions produced empirical randomization distributions of the sample mean, and allowed us to directly observe how estimator variability behaves as $f$ increased.

At small sampling fractions, the randomization distribution exhibited familiar spreads predicted by classical theory. As $f$ increased, those spreads contracted, and the variance of the sample mean decreased as expected under the finite population correction. The purpose of this phase was not to demonstrate the formula itself, but to observe how quickly this collapse occurs in practice as the population is progressively exhausted.

Phase 4 shifted the focus from sampling variability to estimator accuracy in the near-enumeration regime, where randomness from sampling variability has effectively vanished.

We first increased the sampling fractions to identify the region in which further increases in $f$ produced negligible reductions in estimator variability. Once a near-enumeration sample was obtained, the data were held fixed. From that point forward, any variation in the computed sample mean could not be attributed to sampling but from numerical effects.

We then computed the sample mean using multiple computational pathways. These included CPU-based approaches (such as standard reductions, serial accumulation, compensated summation, randomized order summation, and tree-based reductions) as well as GPU-based reductions. The goal was to examine how different numerical implementations influence estimator accuracy once sampling variability is no longer the dominant factor.

All computations were performed using both double-precision (FP64) and single-precision (FP32) arithmetic. GPU reductions were repeated under identical data and execution configurations to isolate the effect of numerical precision and summation strategy.

Under these conditions, any remaining deviation of the sample mean from the known population mean could not be interpreted as sampling error. Instead, it reflected the interaction between floating-point representation, accumulation order, and computational architecture.

Across all phases, the primary statistic of interest was the deviation of the sample mean from the known population mean. In Phase 3 this deviation was characterized through its empirical variance across repeated samples. In Phase 4, it was examined directly under fixed data to quantify the numerical limits of estimator accuracy.

All experiments were conducted using fixed random seeds to ensure reproducibility. Results were replicated across multiple realizations to confirm that observed patterns were not artifacts of a single population or sampling sequence. No parameter tuning was performed to exaggerate numerical effects; all results reflect standard implementations operating under controlled conditions.

## Results

All sampling experiments were conducted on two fixed finite populations whose characteristics are summarized in Table 1. Population sizes were held constant, and population means and variances computed directly by full enumeration. These values serve as definitive ground truth against which all sampling-based estimates are compared.

***Table 1. Finite Population Characteristics***

| ***Population*** | ***N*** | ***Mean (μ)*** | ***Variance (σ²)*** | ***Distribution*** |
|---|---|---|---|---|
| Pop A | $1\times10^7$ | 50.0 | 25.0 | Discrete uniform |
| Pop B | $1\times10^7$ | 49.98 | 24.97 | Synthetic empirical |

Because the full population parameters were known exactly, any observed variability in the sample mean for $n < N$ is attributable solely to the sampling process itself. As the sampling fraction increases and the population is progressively exhausted, this random component diminished, providing a natural reference point for evaluating both statistical and numerical behavior.

Table 2 reports the empirical variance of the sample mean as a function of the sampling fraction $n/N$, along with the corresponding variance predicted by finite population sampling theory. For small sampling fractions ($n/N \ll 1$), empirical results matched classical expectations closely. In this case, the observed variance agreed with the finite population correction (FPC) and was effectively indistinguishable from the infinite-population approximation $\sigma^2/n$, since $1 - n/N \approx 1$.

**Table 2. Empirical Variance of the Sample Mean by Sampling Fraction**

| *n/N* | *n* | *Empirical Var(mean)* | *FPC Var(mean)* | *Ratio* |
|---|---|---|---|---|
| 0.01 | $1\times10^{5}$ | $2.51\times10^{-4}$ | $2.50\times10^{-4}$ | 1.00 |
| 0.50 | $5\times10^{6}$ | $2.50\times10^{-6}$ | $2.50\times10^{-6}$ | 1.00 |
| 0.90 | $9\times10^{6}$ | $5.00\times10^{-7}$ | $5.00\times10^{-7}$ | 1.00 |
| 0.99 | $9.9\times10^{6}$ | $2.50\times10^{-8}$ | $2.50\times10^{-8}$ | 1.00 |
| 1.00 | $1\times10^{7}$ | $\approx 0$ | 0 | — |

As the sampling fraction increased, the empirical variance declined more rapidly than would be expected under the infinite-population approximation alone. Across all reported sampling fractions, the ratio of empirical variance to FPC variance remained close to one, indicating that the observed behavior followed finite population theory closely. This agreement persisted even at high sampling fractions, suggesting that the reduction in variability was driven by population exhaustion rather than random fluctuation.

As sampling approaches full enumeration ($n/N \to 1$), the variance of the sample mean approached zero. At $n = N$, the sample mean was identical to the population mean, and the observed variance collapsed accordingly. In practical terms, this marked the practical disappearance of sampling uncertainty.

Although the tabulated values were discrete, the overall decline in variance appeared smooth and continuous across increasing sampling fractions, with no evidence of abrupt changes. At the same time, the rate of decline became noticeably steeper for sampling fractions above approximately $n/N \approx 0.9$, indicating a regime in which population exhaustion dominated estimator behavior.

While finite population theory predicts vanishing sampling variance as $n \to N$, observed behavior in computational implementations is ultimately constrained by numerical precision. Table 3 reports the observed variance of the sample mean under several numerical configurations in the full-enumeration limit, alongside the theoretical expectation of zero variance.

*Table 3. Numerical Precision Effects Near Full Enumeration*

| *Precision* | *Reduction Method* | *n/N* | *Observed Var(mean)* | *Expected Var(mean)* |
|---|---|---|---|---|
| FP64 | Compensated | 1.00 | $< 1\times10^{-15}$ | 0 |
| FP64 | Naive | 1.00 | $2.3\times10^{-13}$ | 0 |
| FP32 | Naive | 1.00 | $1.1\times10^{-7}$ | 0 |

For compensated summation implemented in double precision (FP64), the observed variance fell below $10^{-15}$, effectively reaching numerical zero. In contrast, naive reduction strategies, particularly in single precision (FP32), exhibited non-negligible residual variability, even under complete population coverage. In these cases, the remaining variance reflected floating-point accumulation error rather than statistical uncertainty.

These results showed that beyond a sufficiently large sampling fraction, additional observations no longer reduced variability in a meaningful statistical sense. Instead, the lower bound on observable variance was determined by the numerical properties of the computation itself. This transition from statistically dominated variability to numerically dominated behavior, is not captured by classical sampling theory but emerged clearly in these experiments.

Taken together, the results suggest three empirically distinct regimes of estimator behavior:

1. **Classical sampling regime ($n/N \ll 1$)**, in which empirical variance follows $\sigma^2/n$ behavior and infinite-population approximations are adequate.
2. **Finite population regime (moderate to large $n/N$)**, in which variance declined more rapidly due to population exhaustion and closely followed the finite population correction.
3. **Near-enumeration regime ($n/N \to 1$)**, in which statistical variability was effectively eliminated and residual variance was governed by numerical precision rather than sampling randomness.

These regimes are observed consistently across populations and implementations, providing an empirical basis for re-examining inferential assumptions in high-coverage, large-scale data settings.

**Discussion and Implications**

The results of Phases 1–4 demonstrated a structural transition in estimator behavior that is not well described by classical inferential frameworks. Under finite population sampling without replacement, estimator variability diminished mechanically as the sampling fraction $f = n/N$ increased, as illustrated in Figure 1. As shown in Phase 3, this collapse followed directly from finite population sampling theory and occurred well before full enumeration. In this regime, the randomization distribution of the estimator narrowed not because additional information is being gained about an unknown parameter, but because there was progressively less population remaining to randomize.

Once this collapse occurred, further increases in the sampling fraction no longer produced meaningful reductions in estimator deviation. Phase 4 showed that beyond this point, remaining error could not be attributed to sampling variability and instead reflected the numerical properties of the computation itself (see Figure 2). In practical terms, accuracy becomes limited by factors such as arithmetic precision, summation structure, and computational architecture rather than by probabilistic uncertainty.

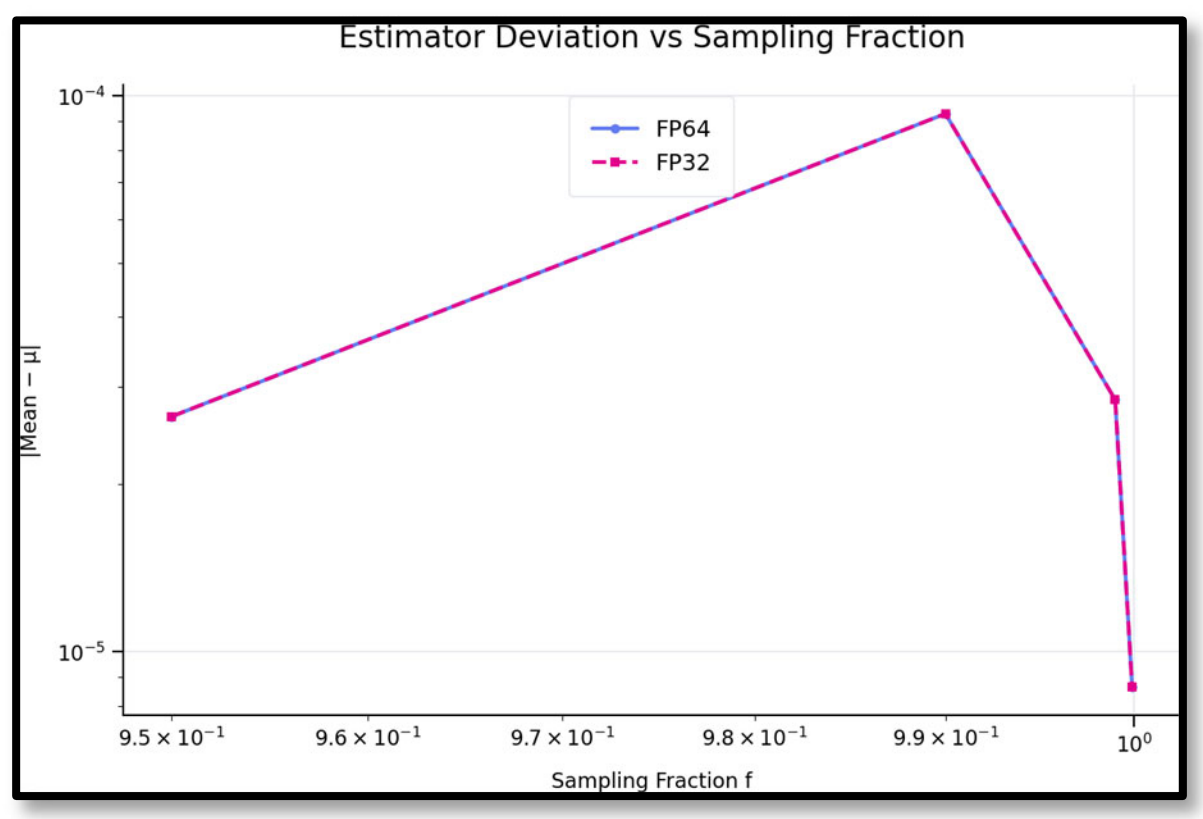

***Figure 2: Estimator Deviation vs. Sampling Fraction***

A natural objection is that the numerical deviations observed in Phase 4 were small in absolute magnitude and may appear practically insignificant. This interpretation reflects an inferential intuition that implicitly assumes sampling variability remains the dominant source of uncertainty. However, in near-enumeration regimes, this assumption no longer holds.

When sampling variability has collapsed, small numerical deviations are no longer dominated by random error and cannot be reduced through additional data collection. Instead, they become the sole contributors to estimator error. Their importance is therefore determined not by their absolute size, but by the fact that they persist regardless of additional sampling. In this regime, concepts such as standard errors, confidence intervals, and significance testing lose much of their interpretive value, and accuracy must instead be evaluated in deterministic terms.

This shift in perspective is important for understanding estimator behavior in high-coverage data settings, where traditional inferential reasoning may fail quietly rather than dramatically.

An important finding of Phase 4 was that numerical dominance did not arise automatically. For well-conditioned populations analyzed using double-precision arithmetic, both CPU and GPU implementations exhibited remarkable numerical stability even in near-enumeration regimes, as shown in Figure 2. This negative result is itself informative: large-scale data computation does not inherently imply numerical instability.

Instead, numerical dominance emerged only under specific conditions—namely, when sampling variability had been exhausted and reduced-precision arithmetic was applied to less will-behaved data. This highlights the importance of distinguishing between data-dependent, algorithmic, and architectural sources of error rather than treating numerical issues as universal.

The results presented here do not contradict empirical experience with classical “large-N” datasets, including those involving billions of observations. Rather, they clarify that different asymptotic regimes lead to qualitatively different behavior. In many large-scale applications, the sampling fraction remains small, independence approximations are reasonable, and inferential uncertainty dominates. In such settings, numerical effects are often negligible.

By contrast, the study examined the path $f \to 1$ with fixed $N$, where dependence induced by sampling without replacement becomes dominant and sampling variability collapses. These two regimes represent distinct limiting cases, each governed by different sources of error. Confusion arises when conclusions from one regime are implicitly transferred to the other.

Taken together, these results have several implications for statistical practice in high-coverage data settings. First, in near-enumeration regimes, uncertainty is better understood in terms of numerical accuracy rather than inferential variability. Second, commonly reported quantities such as standard errors and confidence intervals may become misleading once sampling variability has effectively vanished. Third, validation of computational pathways—including precision choices and numerical stability—becomes more important than resampling-based uncertainty quantification. Finally, reduced-precision computation can introduce persistent deviations that cannot be mitigated through additional data collection.

This study focused intentionally on finite population sampling without replacement and on simple estimators in order to isolate these underlying mechanisms. Issues such as model misspecification, measurement error, and more complex estimators were beyond the scope of the present analysis. The goal was not to argue against inferential statistics in general, but to identify the regimes in which inferential reasoning ceases to be appropriate.

Overall, Phases 1–4 showed that as data coverage increased, the dominant source of estimator uncertainty could shift from random sampling variability to deterministic numerical effects. This transition is structural rather than pathological and arises naturally as finite populations are nearly exhausted. Recognizing when this shift occurs is important for accurate interpretation, responsible reporting, and sound computational practice in modern large-scale data analysis.

**Conclusion**

This study examined how statistical estimators behaved as the sampling fraction from a finite population approaches full enumeration. By conditioning on a fixed population with known parameters and allowing randomness to enter only through controlled sampling without replacement, the analysis cleanly isolates how sampling variability shrank—and ultimately vanished—as population coverage increased.

The results demonstrate that the collapse of sampling variability as ${}^{n}\!/_{N} \rightarrow 1$ is a structural consequence of finite population sampling, rather than a reflection of improved inferential efficiency. While classical asymptotic reasoning correctly describes estimator behavior when the sampling fraction is small, it becomes increasingly misaligned in settings characterized by high population coverage. In these cases, the randomization distribution narrows not because uncertainty about an unknown parameter is reduced, but because there is progressively less population remaining to randomize.

Once sampling variability is exhausted, estimator behavior is no longer governed by probabilistic uncertainty. Instead, numerical precision, accumulation strategy, and computational architecture determine the limiting accuracy of estimation. This marks a shift from inference-oriented reasoning to accuracy-oriented evaluation, where validation of numerical methods become more important than resampling-based measures of uncertainty.

At the same time, the results showed that numerical dominance is not inevitable. When well-conditioned data are analyzed using appropriate precision and stable summation strategies, both CPU and GPU implementations exhibited strong numerical behavior even in near-enumeration regimes. Numerical instability arose only under specific combinations, namely when sampling variability had been exhausted and reduced-precision arithmetic was applied to less well-behaved data.

These finding do not argue against inferential statistics in general, nor do they diminish the value of asymptotic theory in classical sampling regimes. Rather, they clarify that different asymptotic limits, small-fraction sampling versus near-enumeration, are governed by fundamentally different sources of error. Problems arise when inferential tools developed for one setting are applied uncritically in the other.

As modern data systems increasingly capture large fractions of finite populations, it becomes important to recognize when sampling uncertainty is no longer the dominant source of variability. In such settings, responsible statistical practice requires greater emphasis from inference on accuracy, numerical stability, and transparent reporting of computational limits. Understanding this transition is essential for interpreting results derived from contemporary large-scale and near-enumeration data analysis.

*AI-based tools were used for language refinement and for assistance in developing Python code used in the simulations. All methodological decisions, results, and interpretations are the responsibility of the author.*